\documentclass[aps,amsfonts,superscriptaddress, nofootinbib, floatfix, 10pt,twocolumn,prd,groupedaddress]{revtex4-1}
\usepackage{graphicx}
\usepackage{amsmath}
\usepackage[latin1]{inputenc}
\usepackage{amsbsy}
\usepackage{color}
\usepackage{epsfig}
\usepackage{graphicx}
\usepackage{amsmath}
\usepackage{amssymb}
\usepackage{bbm}
\usepackage{amsbsy}
\usepackage{slashed}
\DeclareMathOperator{\tr}{tr}

\usepackage{xcolor}

\usepackage[normalem]{ulem}  

\begin{document}
\title{Unstable quasiparticles as a source of thermodynamic instabilities in the thermal nonlocal Nambu-Jona-Lasinio model.}
\author{F. Marquez}
\affiliation{Facultad de F\'isica, Pontificia Universidad Cat\'olica de Chile, Casilla 306, Santiago 22, Chile.}

\begin{abstract}
It has already been shown that in thermal nonlocal Nambu-Jona-Lasinio models some unphysical behavior, such as a negative pressure, may arise. In this article, it is shown how this behavior can be related to the pressence of highly unstable poles of the propagator of the model, for both the Gaussian and Lorentzian regulators cases. Computations are carried out within the real time formalism, which allows to isolate the contributions from different poles and identify the source of these instabilities. It has also been shown in recent articles how these instabilities are softened by the inclusion of the Polyakov loop when a Gaussian regulator is considered. This article shows how the softening of instabilities can be understood by studying the effect of the Polyakov loop on the poles of the propagator for the Gaussian regulator.
\end{abstract}

\maketitle

\section{Introduction}

There are several aspects of QCD, like the mechanism behind quark confinement or the phase diagram, that are not yet fully understood. Several models have been developed in order to describe some of these properties, such as the bag model \cite{Yo01, Loewe01, Loewe02, Villavicencio01}, Dyson-Schwinger models \cite{Roberts01, Roberts02, Blaschke01} and the linear sigma model \cite{Detar01, Delbourgo01, Lenaghan01, Renato01}. Another attempt in this direction is the Nambu-Jona-Lasinio (NJL) model \cite{Nambu01, Nambu02}. Although it was originally proposed as a model of interacting nucleons, the NJL model is now interpreted as model of interacting quarks and is vastly used to study thermal properties of QCD \cite{Loewe03, Blaschke02, Blaschke03, Buballa01, Klevansky01}.\\

The nonlocal Nambu-Jona-Lasinio (nNJL) model is a generalization of the NJL model \cite{Birse01, Birse02} with a nonlocal interaction that is modulated by a regulator. The nNJL model is also used to study thermal properties of QCD \cite{Loewe04, Scoccola01, Scoccola02, Scoccola03, Weise01}. Thermal computations in nNJL models are usually carried out using the imaginary time formalism. Although calculations are usually simpler in the imaginary time formalism, the physical interpretation of the propagator can be a bit cumbersome. However, in the real time formalism the quark propagator has the usual structure where we have singularities that can be interpreted as quasiparticles. This structure of the propagator allows us to manipulate expresions in order to study all, or only a few of the quasiparticles contributions. Expresions in the real time formalism will usually have the form of a zero temperature contribution plus several terms from thermal contributions, each of them corresponding to a different pole of the propagator. This allows us to track different poles by isolating their contributions to some thermal quantity like the chiral condensate.\\

In a few recent articles \cite{Blaschke04, Blaschke05}, it has been shown that some unphysical instabilities arise within these models which, however, are softened by the inclusion of the Polyakov loop in the case of the Gaussian regulator. This article shows that these instabilities arise, for both Gaussian and Lorentzian regulators, because one is considering highly unstable poles of the propagator. The physical input from the quasiparticle interpretation for the poles of the propagator, allows to comment on the reasons behind these instabilities and their relation with unstable poles. By working in the real time formalism, contributions from different poles can be isolated. In this manner it is possible to isolate the contributions of different quasiparticles and study their relation to the appearance of thermal instabilities. The Polyakov loop is then included in the model for the Gaussian regulator case. By doing so, one is also incorporating new singularities into the propagator. Then, the behavior of these new singularities is studied and they are related to the softening of instabilities. The most commonly used regulators: the Gaussian and the Lorentzian regulators, are considered in this article. Thermal instabilities are present and they can be related to the appearance of unstable poles. These thermal instabilities can be removed in some cases by a careful selection of poles or by a different choice in the parameters of the model.\\ 

The paper is organized as follows. In Sec. II, the nNJL model is introduced and the real time formalism is developed in a general manner. In Sec. III the formalism is applied to the Gaussian regulator. In Sec. IV the formalism is applied to an integer Lorentzian regulator and in Sec. V to a fractional Lorentzian regulator. In Sec. VI a few remarks on how these instabilities may be handled, and the physical motivation for doing so, are made. In Sec. VII a brief discussion on the inclusion of the Polyakov loop in the model, for the Gaussian regulator case, and how this affects the quasiparticles behavior and the occurrence of thermal instabilities is presented. In Sec. VIII conclusions are presented.\\

\section{\lowercase{n}NJL Model in real time formalism.}

The nNJL model is described through the Euclidean Lagrangian
\begin{equation}\mathcal{L}_E=\left[\bar{\psi}(x)(-i\slashed{\partial}+m)\psi(x)-\frac{G}{2}j_a(x)j_a(x)\right],\end{equation}
with $\psi(x)$ being a quark field. The nonlocal aspects of the model are incorporated through the nonlocal currents $j_a(x)$
\begin{equation}j_a(x)=\int d^4y\,d^4z\,r(y-x)r(z-x)\bar{\psi}(x)\Gamma_a\psi(z),\end{equation} 
where $\Gamma_a=(1,i\gamma^5\vec{\tau})$. A bosonization procedure can be performed by defining scalar ($\sigma$) and pseudoscalar ($\vec{\pi}$) fields. Then, in the mean field approximation,
\begin{eqnarray}
\sigma&=&\bar{\sigma}+\delta\sigma\\
\vec{\pi}&=&\delta\vec{\pi},
\end{eqnarray}
where $\bar{\sigma}$ is the vacuum expectation value of the scalar field. In this manner, $\bar{\sigma}$ is closely related to the quiral condensate, and they must behave in exactly the same manner. Also, it was assumed for the pseudoscalar field to have a null vacuum expectation value because of isospin symmetry. Quark fields can then be integrated out of the model \cite{Scoccola02, Scoccola04} and the mean field effective action can be obtained
\begin{equation}\Gamma^{MF}=V_4\left[\frac{\bar{\sigma}^2}{2G}-2N_c\int\frac{d^4q_E}{(2\pi)^4}\tr\ln S_E^{-1}(q_E)\right],\end{equation}
with $S_E(q_E)$ being the Euclidean effective propagator
\begin{equation}S_E=\frac{-\slashed{q}_E+\Sigma(q_E^2)}{q_E^2+\Sigma^2(q_E^2)}.\label{Euprop}\end{equation}
Here, $\Sigma(q_E^2)$ is the constituent quark mass
\begin{equation}\Sigma(q_E^2)=m+\bar{\sigma}r^2(q_E^2).\end{equation}
Finite temperature ($T$) effects can be incorporated through the Matsubara formalism. To do so, one can make the following substitutions
\begin{eqnarray}
V_4&\rightarrow&V/T\\
q_4&\rightarrow&-q_n\\
\int\frac{dq_4}{2\pi}&\rightarrow&T\sum_n,
\end{eqnarray}
where $q_n$ includes the Matsubara frequencies
\begin{equation}q_n\equiv(2n+1)\pi T.\end{equation}
With this, the propagator in Eq. (\ref{Euprop}) will now look like
\begin{equation}S_E(q_n,\boldsymbol{q},T)=\frac{\gamma^4q_n-\boldsymbol{\gamma}\cdot\boldsymbol{q}+\Sigma(q_n,\boldsymbol{q})}{q_n^2+\boldsymbol{q}^2+\Sigma^2(q_n,\boldsymbol{q})}.\label{Euprop2}\end{equation}
It is worth noting that the propagator in Eq. (\ref{Euprop2}) has no singularities. Since there are no poles at some $p^2$, the deffinition of an effective mass for the particle with such propagator is non trivial. Because of this it is harder to physically understand and interpret some quantities in the imaginary time formalism.\\

The $\sigma$ field will evolve with temperature. This evolution can be computed through the grand canonical thermodynamical potential in the mean field approximation $\Omega_{MF}(\bar{\sigma},T,\mu)=(T/V)\Gamma_{MF}(\bar{\sigma},T,\mu)$ \cite{Kapusta01}. Then the value of $\bar{\sigma}$ must be at the minimum of the potential where 
$\partial\Omega_{MF}/\partial\bar{\sigma}=0$, which means
\begin{equation}
\left.\frac{\bar{\sigma}}{G}=2N_cT\sum_n\int\frac{d^3q}{(2\pi)^3}
r^2(q_E^2)\tr 
S_E(q_E)
\right|_{q_4=-q_n}.
\label{gap}
\end{equation}
From this equation one can get the temperature evolution of $\bar{\sigma}$. So far, all of the computations have been made in the imaginary time formalism. Similar derivations are readily available in literature (see for example \cite{Scoccola02, Scoccola06}).\\

The next step is to workout the model in the real time formalism \cite{Loewe04, Yo01, Blaschke04}. In order to do so one should first perform a Wick rotation $q_4=iq_0$ that will take us from Euclidean to Minkowski space. Doing this in Eq. (\ref{Euprop}) will yield the zero temperature Minkowski space propagator
\begin{equation}S_0=i\frac{\slashed{q}+\Sigma(-q^2)}{q^2-\Sigma^2(-q^2)},\label{zp}\end{equation}
where $q^2=-q_E^2$. This propagator has singularities in the complex $q^2$ plane. Each of these singularities may be interpreted as a different quasiparticle of the model. Then, it is possible to define a mass and a decay width for the quasiparticles. If $q^2=\mathcal{M}^2$ is a singularity of the propagator, the following definition can be made
\begin{equation}q^2=\mathcal{M}^2=M^2+iM\Gamma,\label{int}\end{equation} 
where $M$ is the constituent mass of the quasiparticle and $\Gamma$ its decay width. The next step is to obtain the thermal propagator in the real time formalism.\\

In the real time formalism, the number of degrees of freedom is doubled \cite{Ojima01, Ojima02, Kobes01, Landsman01, LeBellac01, Das01}. This means that the thermal propagator is given by a $2\times2$ matrix with elements $S_{ij}$. However, in one-loop calculations only the $S_{11}$ component is necesary. A general expression for $S_{11}$ can be written in terms of the spectral density function (SDF)
\begin{equation}S_{11}=\int\frac{dk_0}{2\pi i}\frac{\rho(k_0,\boldsymbol{q})}{k_0-q_0-i\varepsilon}-n_F(q_0)\rho(q),\label{prop}\end{equation}
where $n_F(q_0)$ is the Fermi-Dirac distribution $n_F(q_0)=({\rm{e}}^{q_0/T}+1)^{-1}$. The SDF can be obtained from 
\begin{equation}\rho(q)=S_+(q)-S_-(q),\end{equation}
where
\begin{equation}S_{\pm}(q)=\pm\oint_{\Gamma^{\pm}}\frac{dz}{2\pi i}\frac{S_0(z\mp i\varepsilon,\boldsymbol{q})}{z-q_0\pm i\varepsilon}.\end{equation}
This is just a generalization of the free particle case where $\rho(q)=S_0(q_0+i\varepsilon,\boldsymbol{q})-S_0(q_0-i\varepsilon,\boldsymbol{q})$. The integration path $\Gamma^{\pm}$ is shown in Fig. 1.\\

\begin{figure}[!htb]
\begin{center}
\includegraphics[scale=0.3]{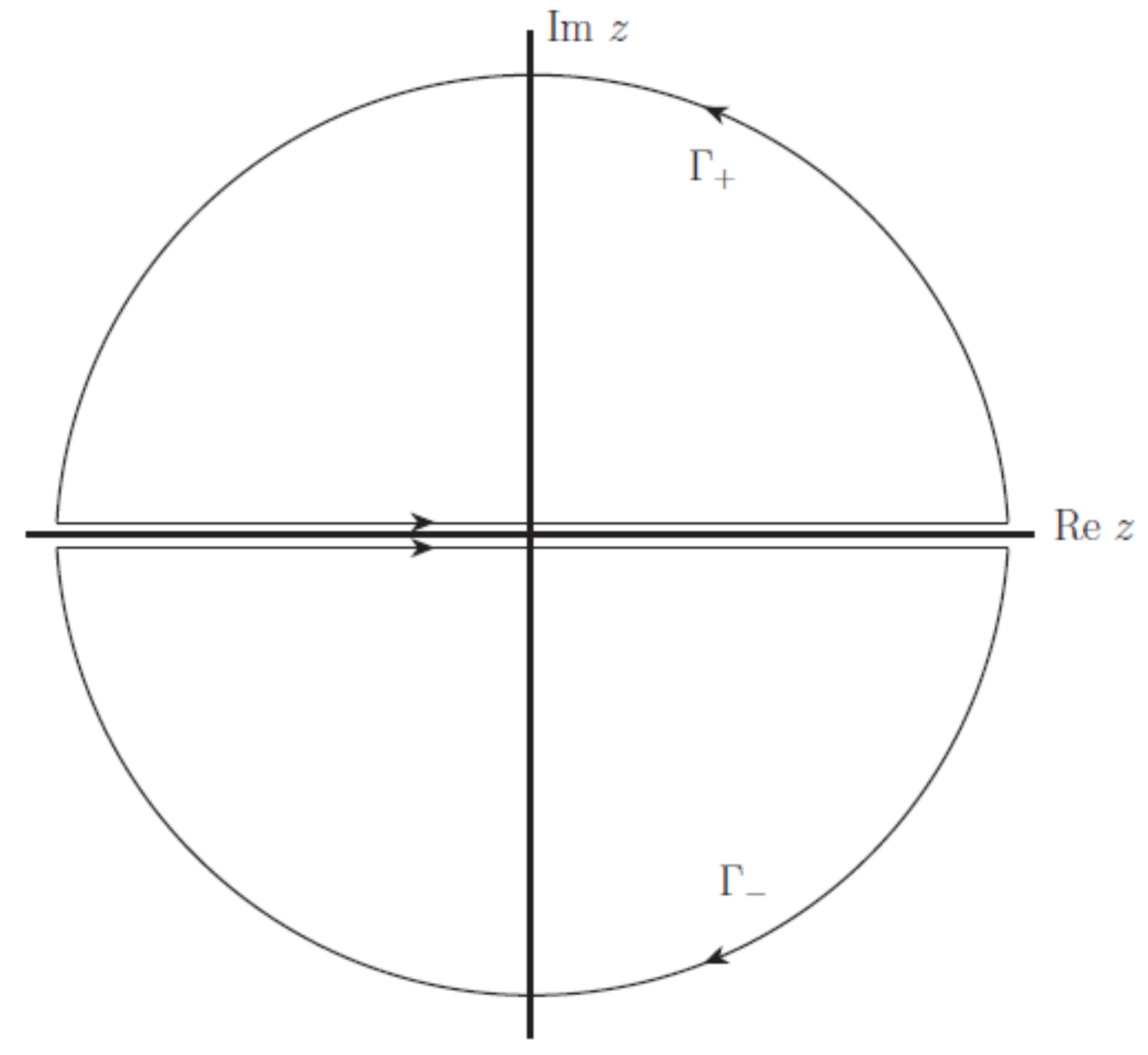}
\caption{Integration path in the definition of $S_{\pm}$.}
\label{pathS}
\end{center}
\end{figure}

The integrations can be performed and the SDF can be found to be 
\begin{equation}\rho(q)=\sum_{\mathcal{M}}i\left[\frac{A(\mathcal{M}^2)}{\mathcal{M}^2-q^2}-\frac{A((\mathcal{M}^2)^*)}{(\mathcal{M}^2)^*-q^2}\right]\label{SDF}\end{equation}
where the sum is over the various poles ($\mathcal{M}$) of the propagator and
\begin{multline}A(\mathcal{M}^2)=\frac{Z(\mathcal{M}^2)}{2E}\left(q_0(\slashed{q}+\Sigma(-\mathcal{M}^2))\right.\\\left.-\gamma^0(q^2-\mathcal{M}^2)\right),\end{multline}
with $E^2=\mathcal{M}^2+\boldsymbol{q}^2$ and where
\begin{equation}Z(\mathcal{M}^2)=\left.\left[\frac{\partial}{\partial q^2}\left(q^2-\Sigma^2(-q^2)\right)\right]^{-1}\right|_{q^2=\mathcal{M}^2},\label{Z}\end{equation}
is the renormalization constant. This calculation is fairly general and is valid for any regulator with real or complex poles of first order. It is worth noting that in Eq. (\ref{SDF}) the contributions from each pole to the SDF are decoupled from each other. This structure is crucial since it allows to isolate the contribution from each pole. The real time thermal propagator can then be obtained by putting Eq. (\ref{SDF}) into Eq. (\ref{prop}). It is clear that the propagator will also have the same structure of the SDF in the sense that contributions form different poles are decoupled from each other. Moreover the propagator will also have the zero temperature contribution decoupled from the finite temperature one, i.e.
\begin{equation}S_{11}(q,T,\mu)=S_0(q)+\tilde{S}(q,T).\label{dec}\end{equation}
Of course, in the case where $\Gamma\rightarrow0$, i.e. when considering real poles, then the propagator reduces to the usual Dolan-Jackiw propagator \cite{Dolan01}
\begin{multline}S_{DJ}(q, M)=(\slashed{q}+M)\left[\frac{i}{q^2-M^2+i\varepsilon}\right.\\\left.-2\pi N(q_0)\delta(q^2-M^2)\right]\end{multline}
The gap equation in the real time formalism should now be obtained through the substitution $S_E\rightarrow S_{11}$ in Eq. (\ref{gap}). Using the structure in Eq. (\ref{dec}) yields
\begin{equation}\frac{\partial\Omega_{MF}}{\partial\bar{\sigma}}=g_0(\bar{\sigma})+\tilde{g}(\bar{\sigma},T)=0,\label{potential}\end{equation}
where
\begin{eqnarray}
g_0(\bar{\sigma})=\frac{\bar{\sigma}}{G}-\frac{N_c}{\pi^2}\int_0^\infty dq_Eq_E^3\frac{r^2(q_E^2)\Sigma(q_E^2)}{q_E^2+\Sigma^2(q_E^2)}\\
\tilde{g}(\bar{\sigma},T)=-2N_c\int\frac{d^4q}{(2\pi)^4}r^2(-q^2)\tr\tilde{S}(q,T)\label{gtilde},
\end{eqnarray}
and where again $\tilde{g}(\bar{\sigma},T)$ has all of the finite temperature contribution. By putting the expression for $\tilde{S}$ into Eq. (\ref{gtilde}) we get
\begin{multline}\tilde{g}(\bar{\sigma},T)=2iN_c\sum_{\mathcal{M}}Z(\mathcal{M}^2)\Sigma(-\mathcal{M}^2)\\\times\int\frac{d^4q}{(2\pi)^4}\frac{r^2(-q^2)}{E}2n_F(q_0)\\\times\left[\frac{q_0}{\mathcal{M}^2-q^2}-\frac{q_0}{(\mathcal{M}^2)^*-q^2}\right]+C.\label{g}\end{multline}
The integration in $q_0$ can be performed along the path shown in Fig. 2.

\begin{figure}[!htb]
\begin{center}
\includegraphics[scale=0.4]{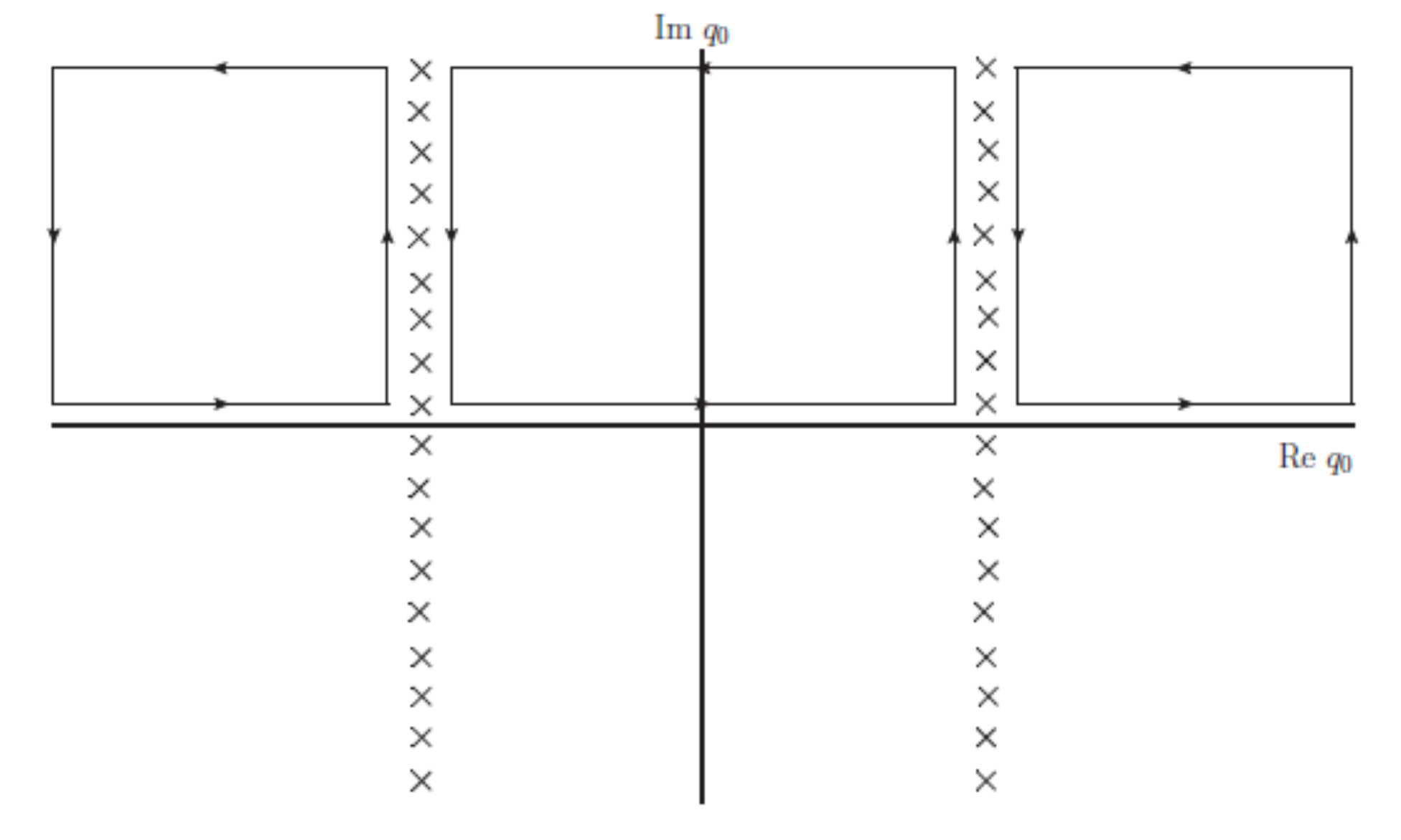}
\caption{Integration path for the thermal part of the gap equation. The poles of the Fermi-Dirac distribution are marked with crosses}
\end{center}
\end{figure}

The integration along the vertical lines where Re $q_0\rightarrow\pm\infty$ gives a divergent contribution that, however, is independent of temperature and will be cancelled by the constant $C$ in Eq. (\ref{g}). The integration then can be computed to give
\begin{multline}\tilde{g}(\bar{\sigma},T)=-\frac{N_c}{\pi^2}\sum_{\mathcal{M}}\left[Z(\mathcal{M}^2)\Sigma(-\mathcal{M}^2)r^2(-\mathcal{M}^2)\right.\\\times\int dkk^2\frac{2n_F(E)}{E}+\left.\left(\mathcal{M}^2\rightarrow(\mathcal{M}^2)^*\right)\right].\label{gfin}\end{multline}

This is the final expression for the gap equation in the real time formalism. Once again, this expression has the contributions from different poles decoupled from each other which is crucial for studying the contributions from different poles independently. Every quasiparticle mass and decay width will evolve with temperature. If we are interested in studying the behaviour of just one quasiparticle by itself, we can toss all of the other terms in Eq. (\ref{gfin}) coming from different poles and solve solely for the quasiparticle we are interested in. Formally, this can be done by deforming the integration path in Fig. \ref{pathS} around the undesired poles in order to exclude them. This will be of the most importance when trying to identify the source of the thermal instabilities that arise in this type of models.\\

The main quantity we are going to look at is $\bar{\sigma}$. It should behave in the same manner as a chiral condensate. This means that one would expect for $\bar{\sigma}$ to monotonically decrease as temperature increases. Otherwise, we would have a condensate that becomes larger at larger temperatures. The gap equation in Eq. (\ref{gfin}) will allow us to get the behavior of $\bar{\sigma}$ as a function of temperature. Any growth of $\bar{\sigma}$ will then be considered an ``instability'' since it does not correspond to the usual behavior of a condensate. \\

\section{Gaussian Regulator.}

Let us now consider the nNJL model with a Gaussian regulator of the form
\begin{equation}r(q^2)={\rm e}^{q^2/\Lambda^2}.\end{equation}
Two different sets of parameters will be considered for this regulator. They are conveniently chosen because of the pole structure the propagator exhibits with such parameters. The parameters are shown in Table I. Set A is taken from reference \cite{Blaschke04} and set B from \cite{Scoccola05}.

\begin{table}[!h]
\begin{tabular}{|c|c|c|c|c|}
\hline
Set&$\Lambda$(MeV)&$m$(MeV)&$G\Lambda^2$&$\bar{\sigma}_0$(MeV)\\
\hline
A&687&6&28.43&677.8\\
B&1042.2&4.6&15.08&235\\
\hline
\end{tabular}
\caption{Both parameters set used for the Gaussian regulator case. $\bar{\sigma}_0$ is the mean value of the scalar field at zero temperature.}
\end{table}

Then, by solving $q^2-\Sigma^2(-q^2)=0$ one can find the poles of the propagator, which means
\begin{eqnarray}
\mbox{Re}(q^2-\Sigma^2(-q^2))=0\label{Re}\\
\mbox{Im}(q^2-\Sigma^2(-q^2))=0.\label{Im}
\end{eqnarray}

\begin{figure}[!h]
\begin{center}
\includegraphics[width=200pt]{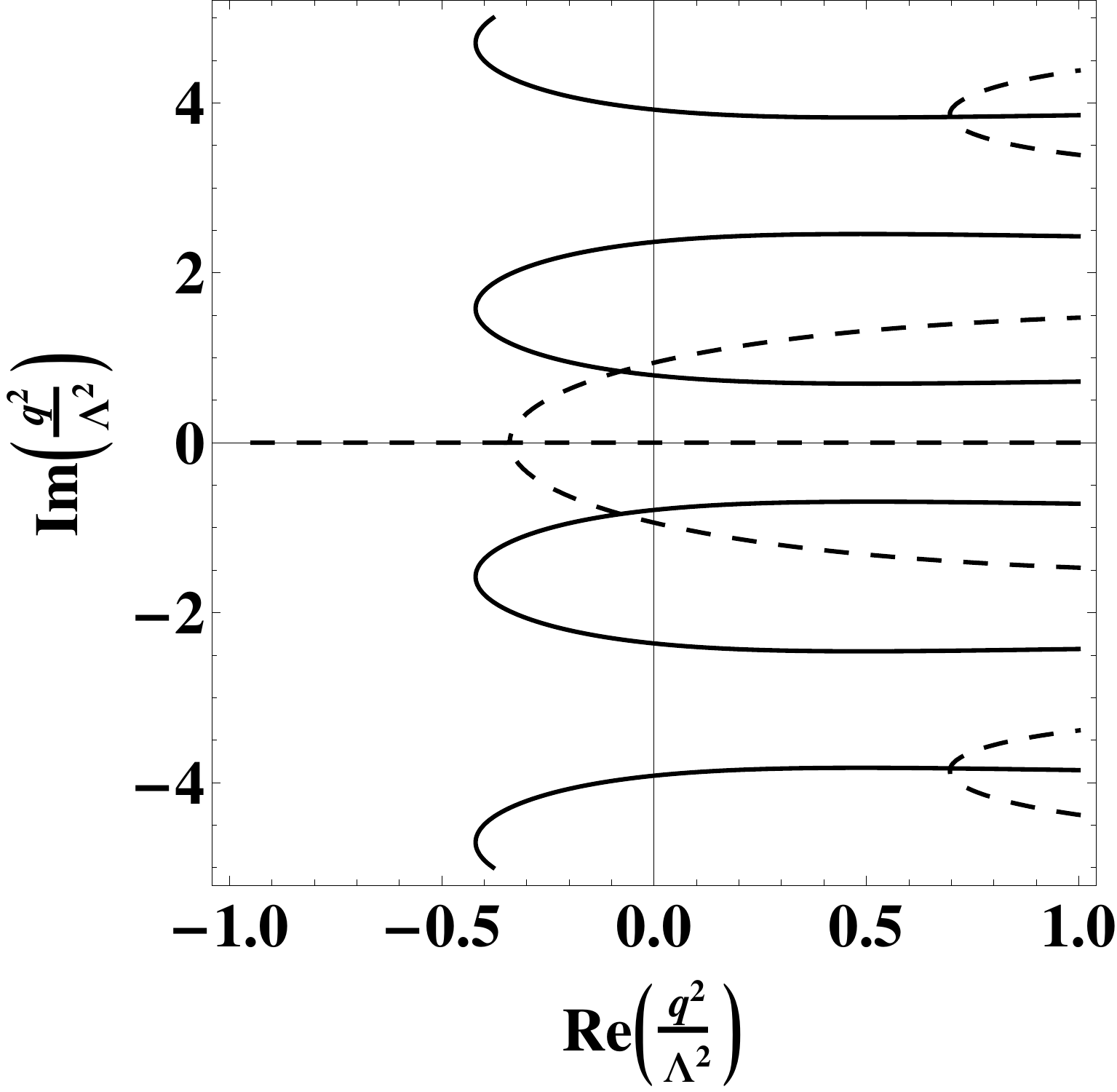}
\end{center}
\caption{Poles of the propagator for set of parameters A for the Gaussian regulator in the $q^2$ plane. The dashed lines are solutions to Eq. (\ref{Im}) and the solid lines solutions to Eq. (\ref{Re}).}
\label{f1a}
\end{figure}

Fig. \ref{f1a} shows solutions to Eqs. (\ref{Re}) and (\ref{Im}) for parameter set A. The poles of the propagator are found at the intersection of the dashed and solid lines. The first pole that appears has a negative real part. If we follow our identification in Eq. (\ref{int}), this means that this pole represents a quasiparticle with negative squared mass. However, one could write $q^2=\left(M+i\frac{\Gamma}{2}\right)^2$, in which case the pole simply has a decay width greater than its mass. We will therefore consider poles like this to be highly unstable poles. It is also important to note that poles of this kind, do not have a clear quasiparticle interpretation. By having such a big decay width they are very short-living states.\\

The propagator will exhibit an infinite number of poles. All of them, except the first pole we already discussed, will be complex poles with positive real parts. However, they will also exhibit big imaginary parts and must be treated as highly unstable poles too.

\begin{figure}[!h]
\begin{center}
\includegraphics[width=200pt]{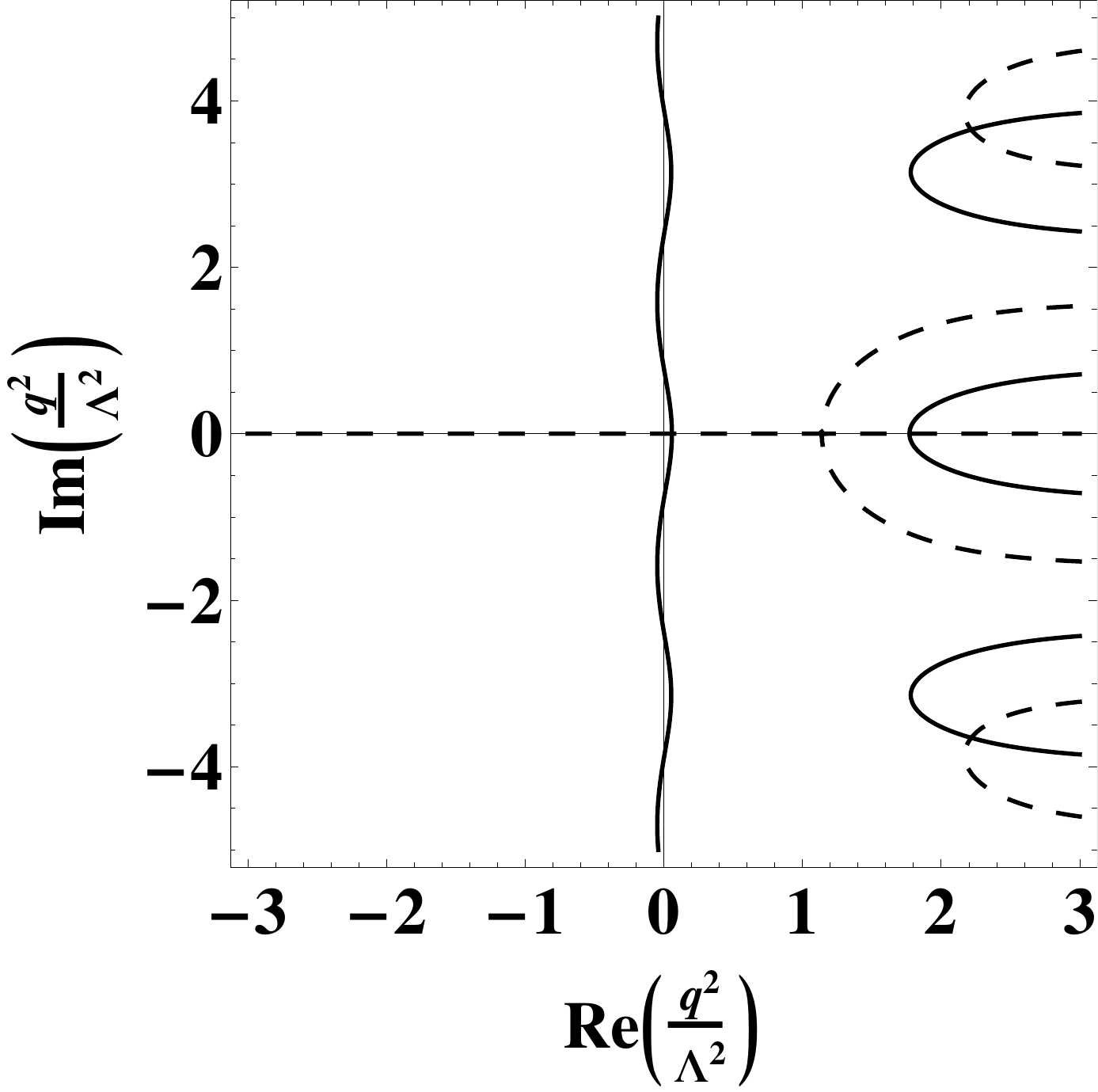}
\end{center}
\caption{Poles of the propagator for set of parameters A for the Gaussian regulator in the $q^2$ plane. The dashed lines are solutions to Eq. (\ref{Im}) and the solid lines solutions to Eq. (\ref{Re}).}
\label{f1b}
\end{figure}

Fig. \ref{f1b} shows solutions to Eqs. (\ref{Re}) and (\ref{Im}) for parameter set B. The poles of the propagator are found at the intersection of the dashed and solid lines. Quite differently from what was found for set of parameters A, in this case, the first two poles of the propagator have vanishing imaginary parts. These states can clearly be interpreted as deconfined quasiparticles \cite{Loewe04}. On the other hand, and similarly to what was found in parameter set A, there are an infinite number of other poles that are complex poles with big imaginary parts, i.e. with decay widths of same order, or greater, than their masses.\\

Our next step is to solve Eq. (\ref{gfin}) in order to get the behavior of $\bar{\sigma}$ as a function of temperature. In the Gaussian regulator case, highly unstable poles are present in our propagator. It is reasonable to expect some odd behavior of $\bar{\sigma}$ in the presence of such poles. On the other hand, a pole with a positive real part and a vanishing imaginary part is a stable quasiparticle and one would expect from it the usual behavior of a condensate. This makes the study of the poles of the propagator a crucial matter in order to understand the behavior of $\bar{\sigma}$. Because of this, before solving Eq. (\ref{gfin}), it is important to take a closer look to the poles of the propagator.\\

\begin{figure}[!htb]
\begin{center}
\includegraphics[width=255pt]{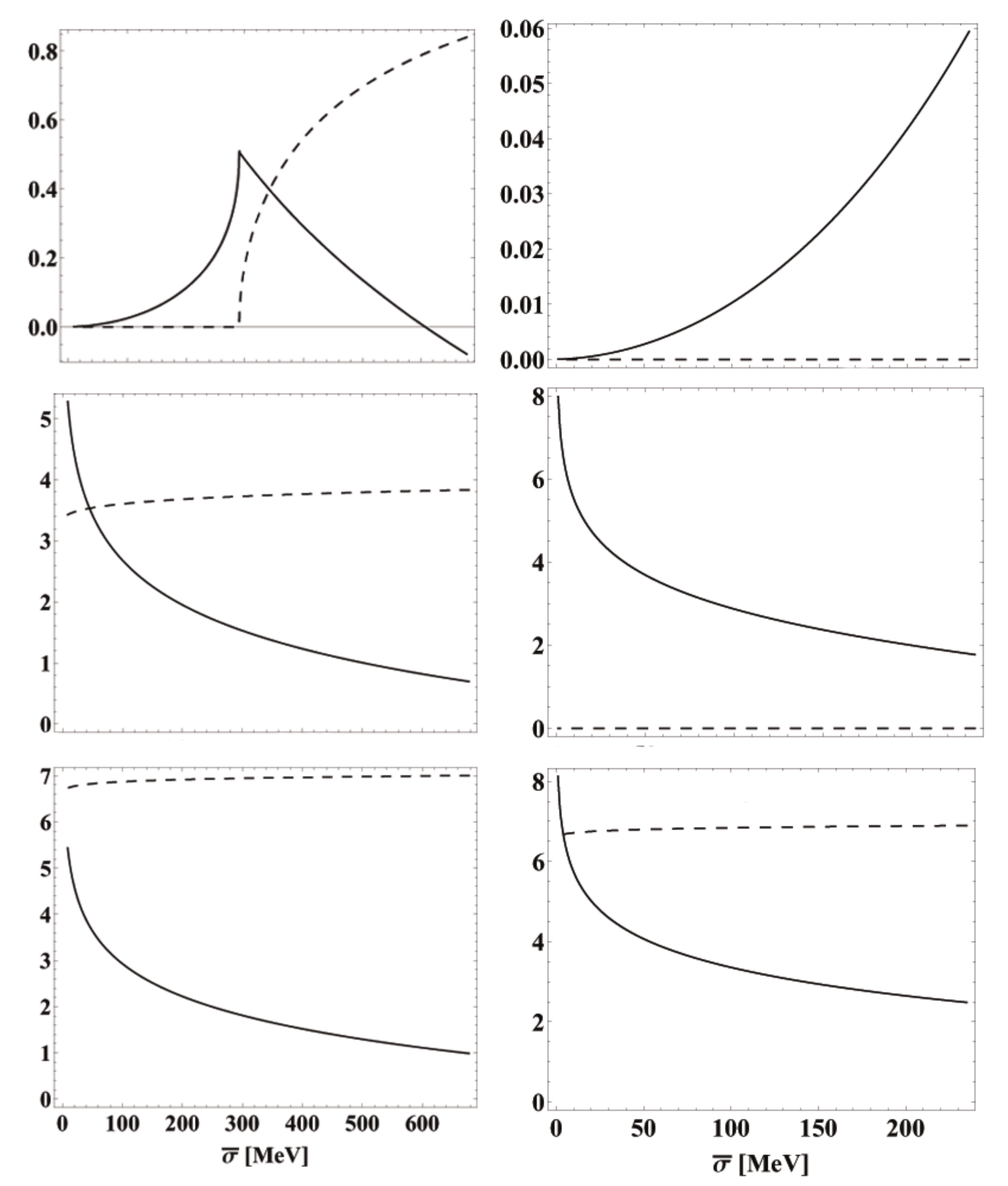}
\end{center}
\caption{Behavior of the first three poles of the propagator, with a Gaussian regulator, as a function of $\bar{\sigma}$ for set A (plots on the left) and set B (plots on the right). The solid lines represent $\mbox{Re}(q^2/\Lambda^2)$ and the dashed lines represent $\mbox{Im}(q^2/\Lambda^2$)}
\label{f2}
\end{figure}

Fig. \ref{f2} shows the behavior of the first three poles of the propagator for both sets as a function of $\bar{\sigma}$. As can be seen, the first pole of set A has a negative real part for $\bar{\sigma}=\bar{\sigma}_0$. However, at lower values, $\bar{\sigma}\approx300$ MeV, the pole has a positive real part and a vanishing imaginary part, i.e. it has turned into a positive real pole and hence, a well defined quasiparticle. On the other hand, for set B, we have two positive real poles and no pole with a negative real part. All of the other poles will turn out to be poles with positive real parts and nonvanishing imaginary parts. They will, however, be much more massive than the poles here considered and should have smaller contributions to the behavior of $\bar{\sigma}$. The gap equation can then be solved considering only a few of the poles and the result will not differ greatly from the complete calculation. Let us now solve Eq. (\ref{gfin}) for these three poles in order to get $\bar{\sigma}$ as a function of temperature for both sets.

\begin{figure}[!htb]
\begin{center}
\includegraphics[scale=0.4]{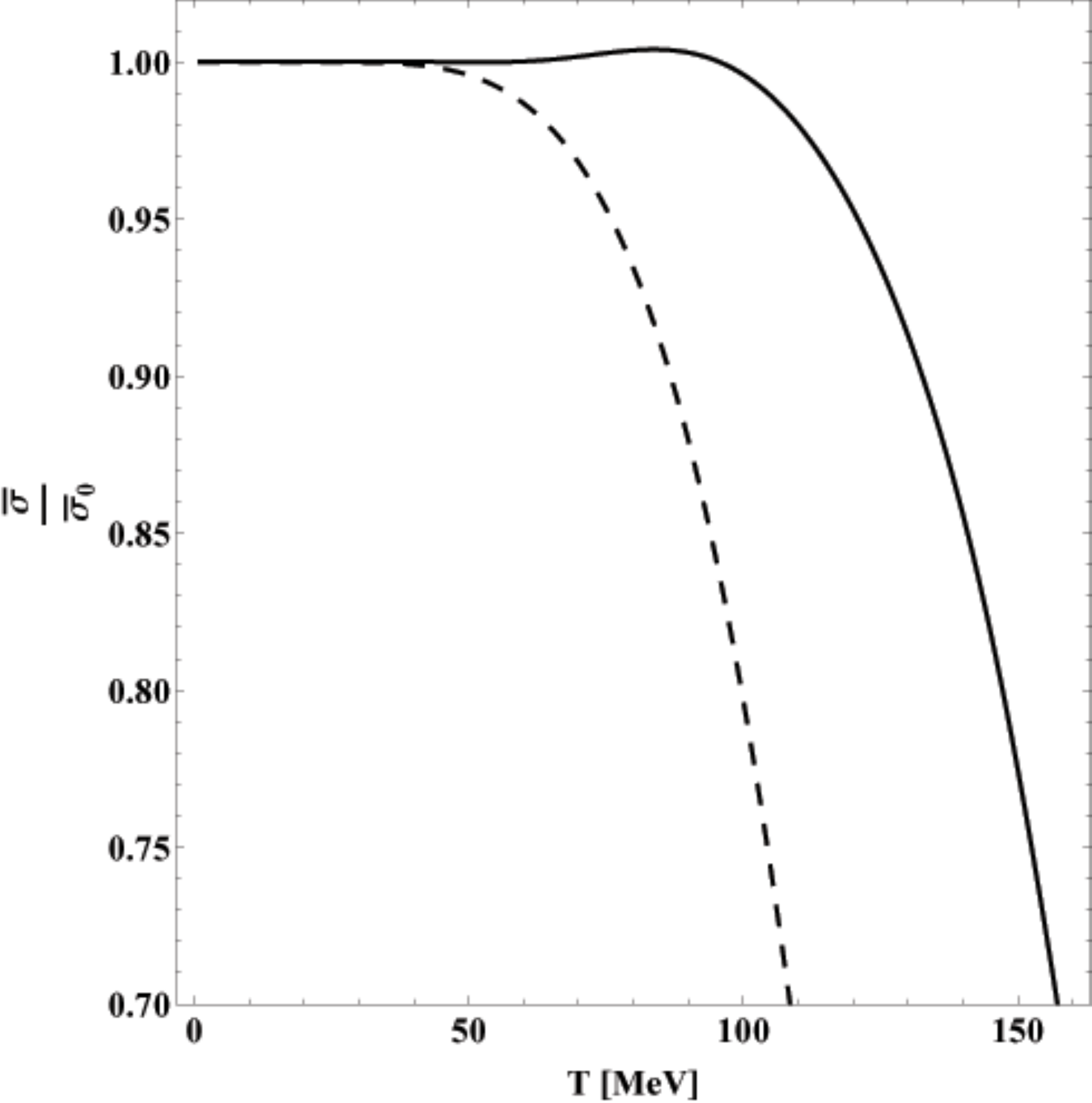}
\end{center}
\caption{Behavior of $\bar{\sigma}$ as a function of temperature. The solid line corresponds to set A and the dashed line to set B. In both cases the the first three poles of the propagator are being considered.}
\label{f3}
\end{figure}

As can be seen from Fig. \ref{f3}, for set A, $\bar{\sigma}$ rises with temperature between $T\approx80-110$ MeV. This is exactly the kind of thermal instability found in \cite{Blaschke04}. However, for set B there is no such instability. Let us recall that the main difference between sets A and B was the presence of a pole with a negative real part in set A as opposed to two positive real poles in set B. One could suspect then that the rising of $\bar{\sigma}$ with temperature in set A is a consequence of the presence of this odd pole. To better understand this, let us solve again Eq. (\ref{gfin}) but now for three different cases: considering the first three poles of set A, considering only the first pole of set A and considering only the second and third poles of set A.

\begin{figure}[!]
\begin{center}
\includegraphics[scale=0.4]{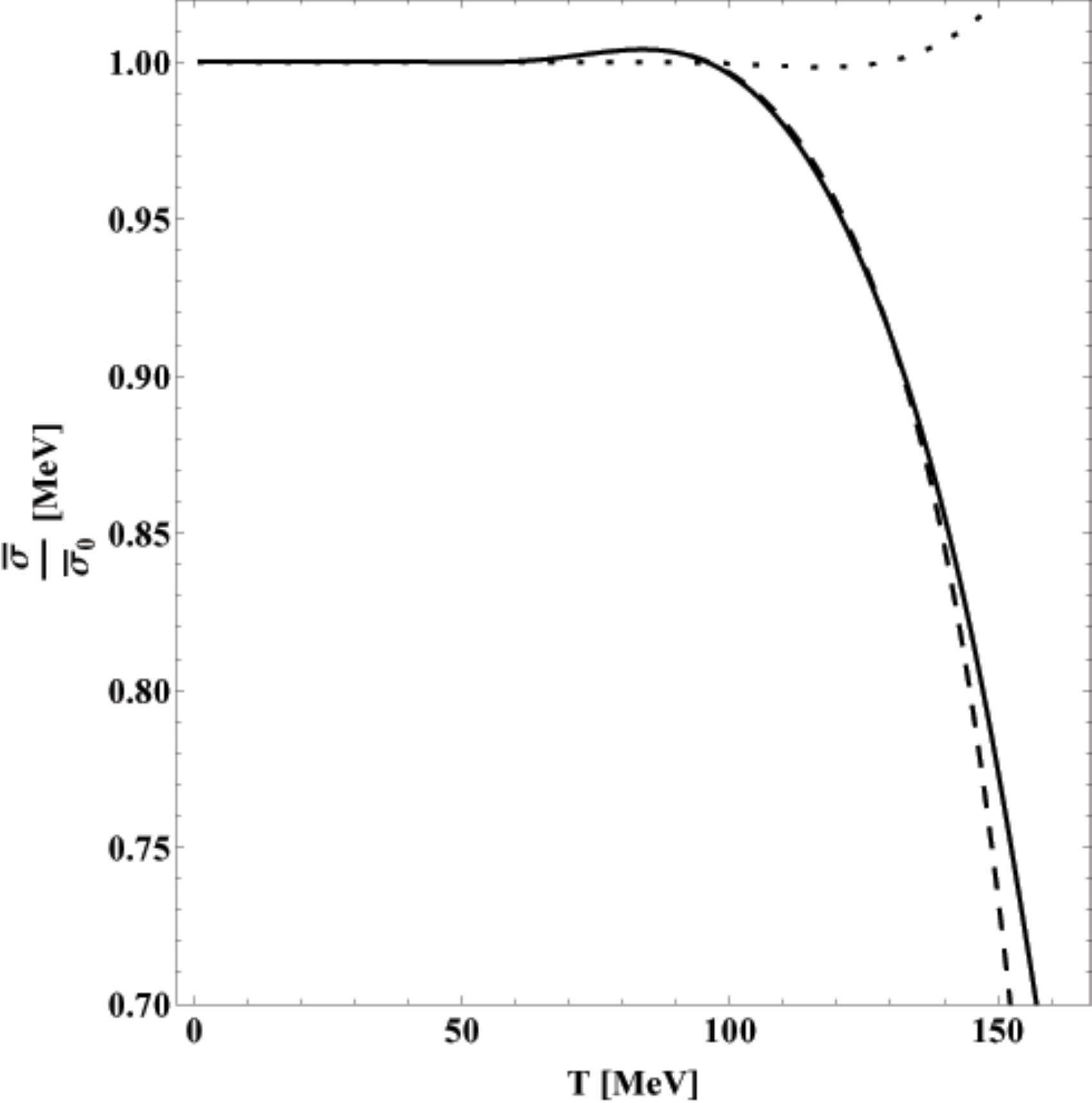}
\end{center}
\caption{Behavior of $\bar{\sigma}$ as a function of temperature. The solid line corresponds to the solution for the first three poles of set A. The dashed is the solution with only the first pole of set A and the dotted line is the solution with only the second and third poles of set A.}
\label{f4}
\end{figure}

As can be seen from Fig. \ref{f4} when we take into account the first pole (solid and dashed lines in Fig. \ref{f4}) the instability remains. When we consider only the second and third poles (dotted line in Fig. \ref{f4}), the initial rising is gone, but a new rising appears now at a higher temperature. This means that the initial instability that was observed when solving with all of the three poles being considered, was caused by the pole with a negative real part. The new instability that appears in the dotted line of Fig. \ref{f4} is caused by the second and third poles. This instability was not present when solving considering all three poles because, at the temperature values at which the instability appears, $\bar{\sigma}$ had already decreased enough so that the first pole had turned to a well defined quasiparticle and, being much less massive than the second and third poles, its contribution was dominant over the others.\\

This analysis has been performed with the Gaussian regulator because it allows to compare the results of this article to those of reference \cite{Blaschke04}, however, the Gaussian regulator is far from ideal to study these situations. On the one hand it has an infinite number of poles which makes the full computation extremely challenging. On the other hand the pole with a negative real part, for set A, is also the only pole that, at lower $\bar{\sigma}$, is a stable pole. This means that the same pole that produces the instability is the only ``well behaved'' pole that will also make our condensate rapidly decrease as temperature increases. In this manner it seems that achieving the usual behavior for $\bar{\sigma}$ when considering the Gaussian regulator with parameter set A may be impossible. Because of this, it is useful to consider other regulators, such as the Lorentzian regulator, that exhibit a finite number of poles and thus allow us to better study how each pole affects the behavior of $\bar{\sigma}$.\\

\section{Integer Lorentzian regulator.}

Let us now consider the following regulator in Minkowski space
\begin{equation}r(q^2)=\frac{1}{1+\left(-\frac{q^2}{\Lambda^2}\right)^2}.\end{equation}
The following parameters are taken from \cite{Scoccola05}: $m=$4.6 MeV, $\bar{\sigma}_0=$216 MeV, $\Lambda=$868 MeV and $G\Lambda^2=$9.61. Eqs. (\ref{Re}) and (\ref{Im}) can be solved to find the poles of the propagator with this regulator

\begin{figure}[!htb]
\begin{center}
\includegraphics[scale=0.4]{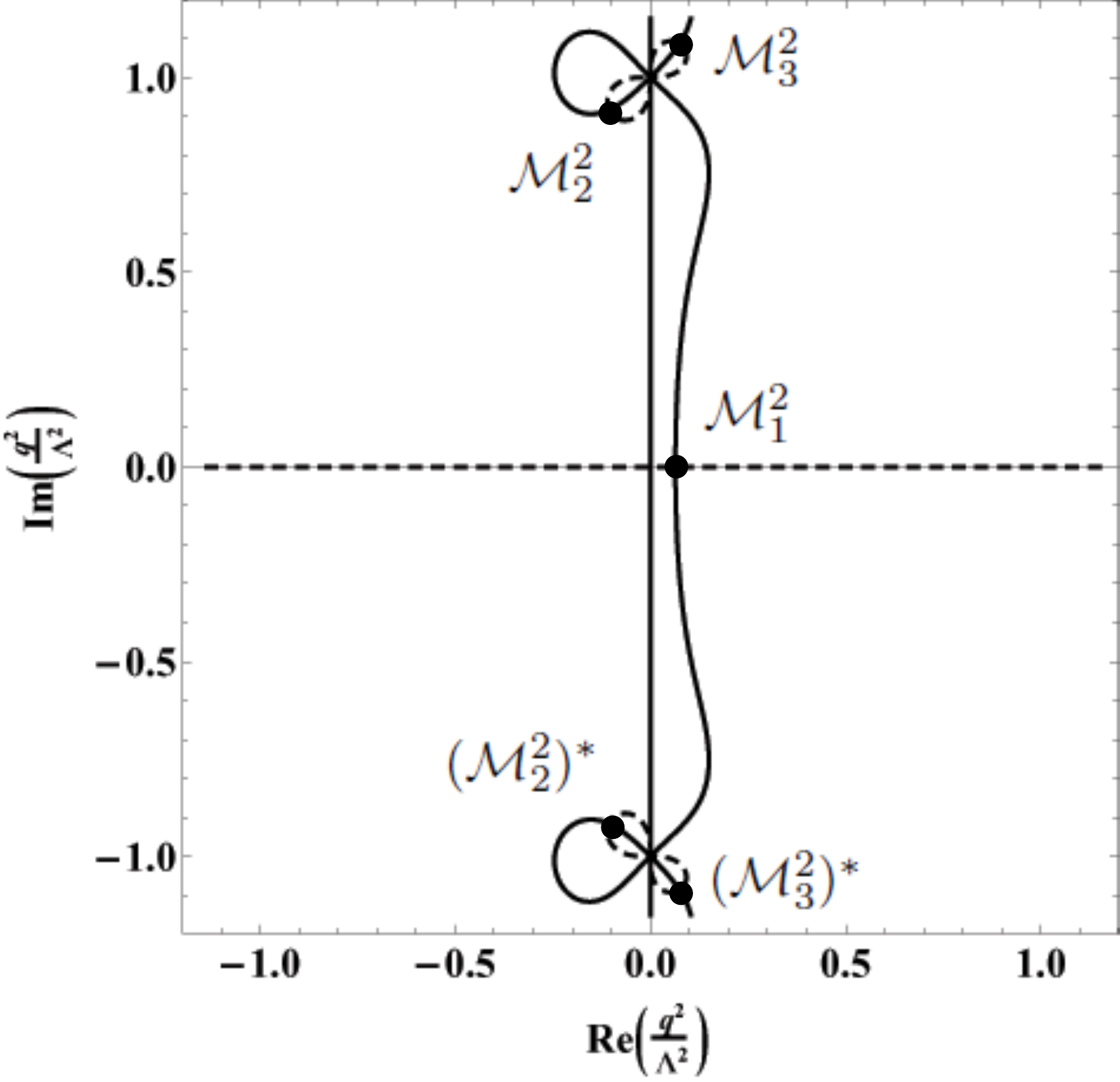}
\end{center}
\caption{Poles of the propagator for the integer Lorentzian regulator. The dashed lines are solutions to Eq. (\ref{Im}) and the solid lines solutions to Eq. (\ref{Re})}
\label{f5}
\end{figure}

As can be seen from Fig. \ref{f5} there are three poles for this regulator. One of the poles is real ($\mathcal{M}_1$) while the other two are complex conjugate pairs, and one of them ($\mathcal{M}^2_2$) has a negative real part. Similarly to what was done in the previous section Eq. (\ref{gfin}) can be solved to get the behaviour of $\bar{\sigma}$ as a function of temperature. It will be done in three different cases: By counting all of the three poles, by counting only $\mathcal{M}_1$ and by counting only $\mathcal{M}_2$.

\begin{figure}[!htb]
\begin{center}
\includegraphics[scale=0.4]{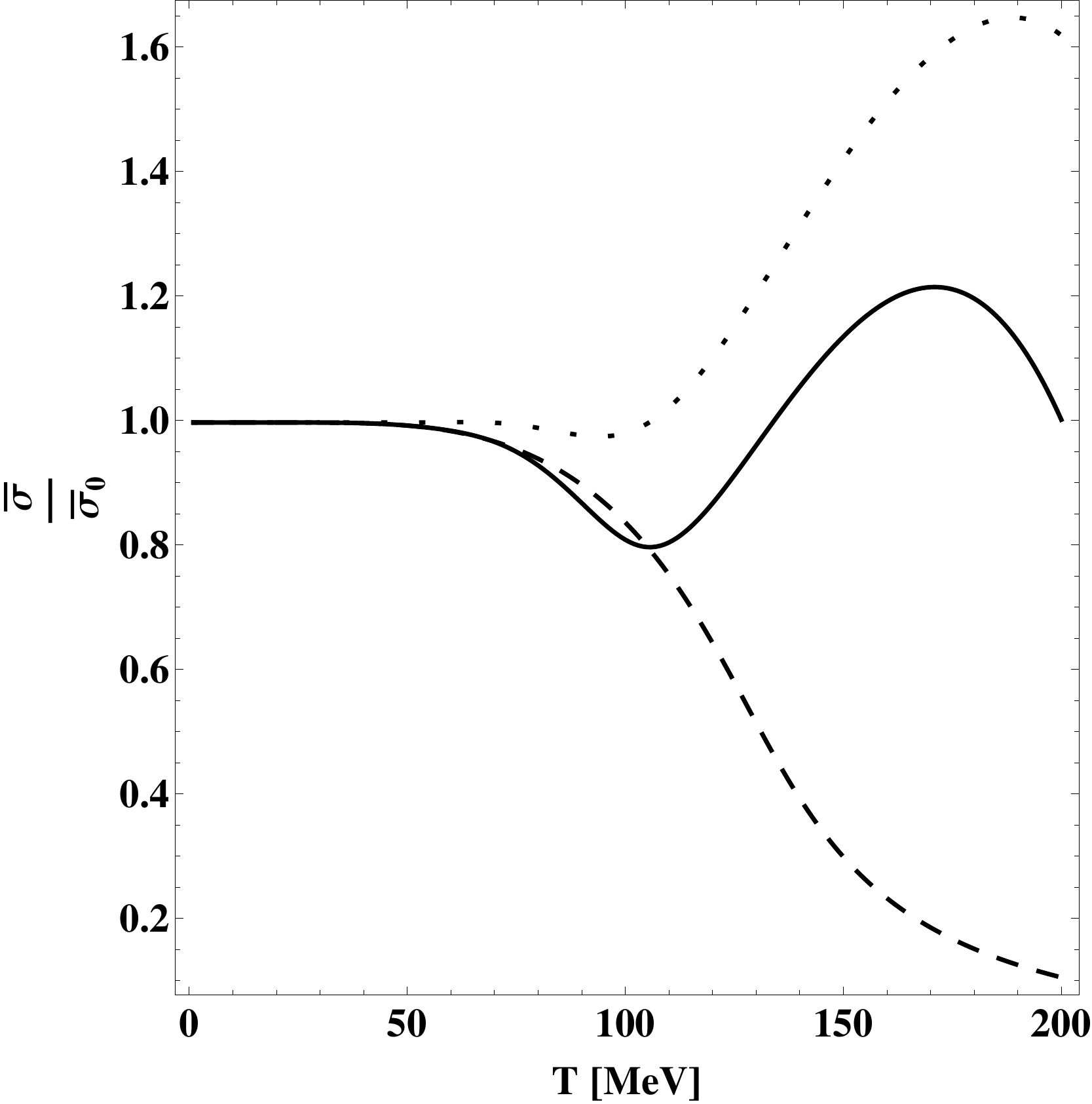}
\end{center}
\caption{Behavior of $\bar{\sigma}$ as a function of temperature. The solid line corresponds is solution from counting all the poles. The dashed line is the solution for counting only $\mathcal{M}_1$ and the dotted line is the solution for counting only $\mathcal{M}_2$ and $\mathcal{M}_3$.}
\label{f6a}
\end{figure}

As can be seen from Fig. \ref{f6a} instabilities appear with this regulator as well. However, if we only consider the real pole (the dashed line in Fig. \ref{f6a}) then we have no instability. The instability arises from the highly unstable poles. If we only consider these poles (dotted line in Fig. \ref{f6a}) then the instabilities become much larger. This is the same type of behavior that was present with the Gaussian regulators. Those poles with big decay widths, produce these type of thermal instabilities. However, if the highly unstable poles are neglected (dashed line in Fig. \ref{f6a}), then the instabilities dissappear.\\ 

Just as in the Gaussian case, it is the real pole that produces no instability. One might expect the same thing from a complex pole with a small imaginary part, i.e. a well defined, confined quasiparticle. However, none of the regulators considered so far exhibits such a pole. This will be the case in the next section.

\section{Fractional Lorentzian regulator.}

Lorentzian regulators with fractional exponents have become interesting because they are able to reproduce lattice data from the light quark propagator \cite{Scoccola04}. Inspired by that we consider the following regulator in Euclidean space
\begin{equation}r(q_E^2)=\frac{1}{1+\left(\frac{q_E^2}{\Lambda^2}\right)^{3/2}}.\end{equation}
On performing the rotation to Minkowski space one has to define how the half integer exponent will be understood \cite{Yo01}. In Minkowski space, momentum may take complex values and so the regulator is a multivalued function of the momentum. If we take $q^2/\Lambda^2\equiv R{\rm e}^{i\theta}$, then we define in Minkowski space
\begin{equation}r(q^2)=\frac{1}{1+R^{3/2}{\rm e}^{\frac{3}{2}i(\theta+\pi)}}.\end{equation}
With this definition the multivalued nature of our regulator is preserved. Therefore there will be two Riemann sheets and hence, poles for our propagator will be found in both of them. This propagator will then have several poles. For our analysis we will work in the chiral limit where $m=0$. In this manner the number of poles is significantly reduced and the model is better suited to study the effects of each pole as is the aim of this article. The following parameters are taken from \cite{Yo01}: $m=0$ MeV, $\bar{\sigma}_0=261$ MeV, $\Lambda=$635 MeV and $G\Lambda^2=$10.81. 

\begin{figure}[!htb]
\begin{center}
\includegraphics[width=220pt]{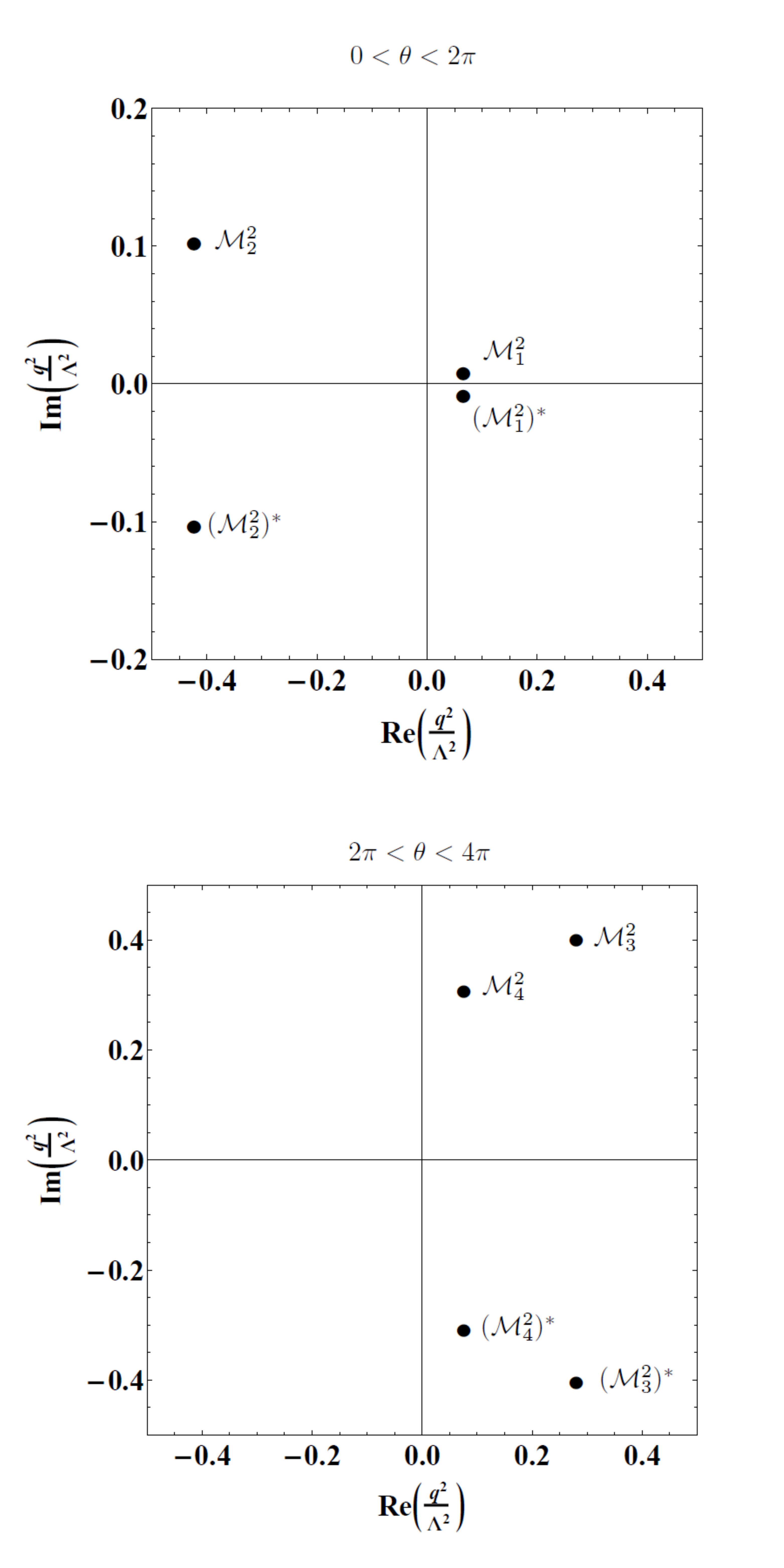}
\end{center}
\caption{Poles of the propagator for the half-integer Lorentzian regulator.}
\label{f8}
\end{figure}

In Fig. \ref{f8} the poles of the propagator are shown. There is one pole with a negative real part in the first sheet ($\mathcal{M}_2$). In the second sheet there are two other poles with big imaginary parts ($\mathcal{M}_3$ and $\mathcal{M}_4$). So far, one would expect some instabilities from these poles. However, in the first sheet, we also have one pole with a positive real part and a small imaginary part ($\mathcal{M}_1$). This is exactly the kind of pole that was not present neither in the Gaussian regulator nor in the integer Lorentzian regulator and one would expect for this pole to yield no instability.

\begin{figure}[!htb]
\begin{center}
\includegraphics[scale=0.4]{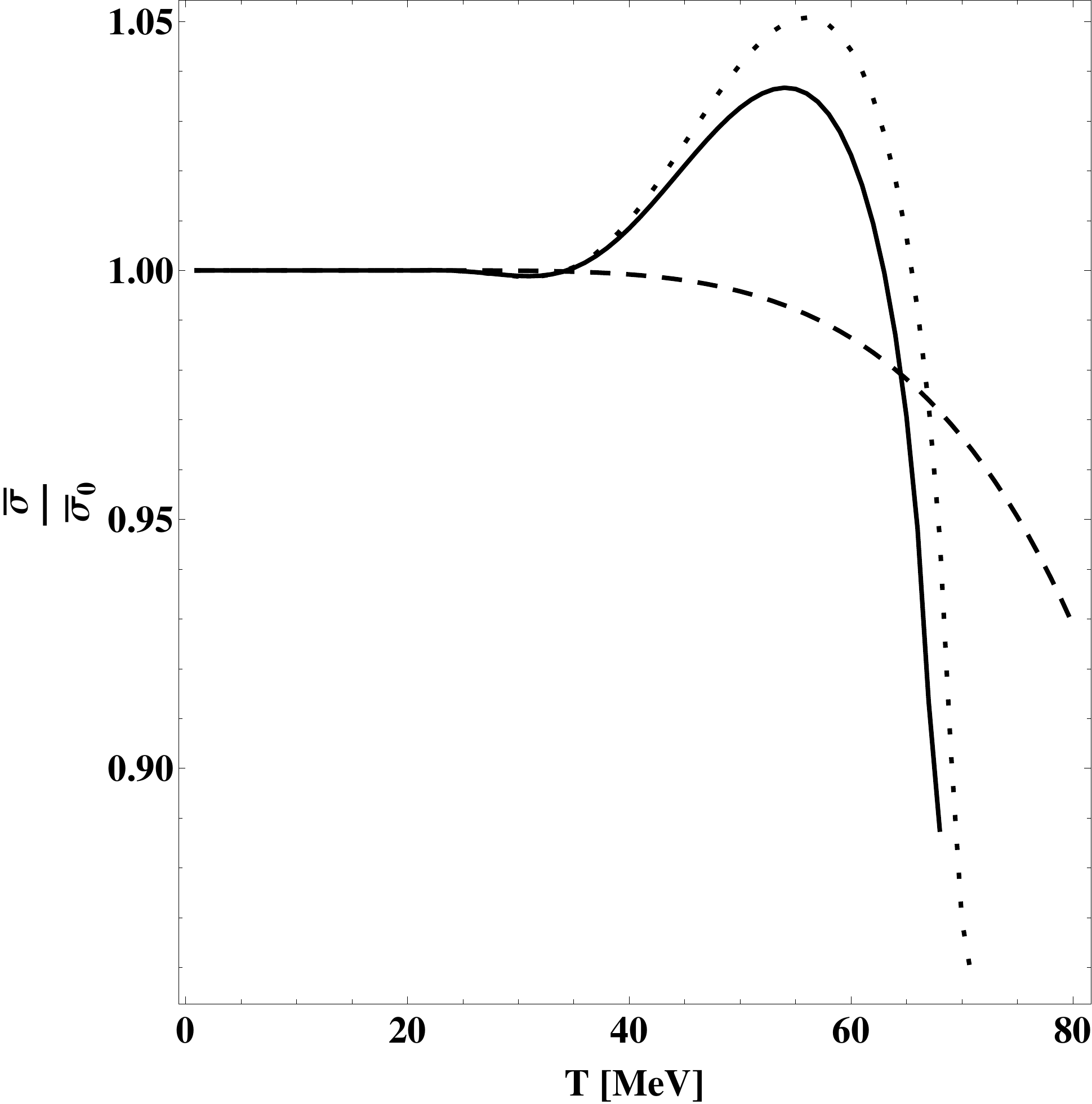}
\end{center}
\caption{Behavior of $\bar{\sigma}$ as a function of temperature. The solid line corresponds is solution from counting all the poles. The dashed line is the solution from counting only $\mathcal{M}_1$ and the dotted line is the solution from counting only $\mathcal{M}_2$, $\mathcal{M}_3$ and $\mathcal{M}_4$.}
\label{f9}
\end{figure}

As can be seen from Fig. \ref{f9} an instability is present when all of the poles of the model are considered. If only the first pole ($\mathcal{M}_1$ in Fig. \ref{f8}) is taken into account, then the instability dissapears. However, if we take into account only the highly unstable poles ($\mathcal{M}_2$, $\mathcal{M}_3$ and $\mathcal{M}_4$ in Fig. \ref{f8}), then we can see that the the instability increases. This reinforces the statement that such instabilities arise from considering condensates of highly unstable quasiparticles.\\

\section{Handling the instabilities.}

So far, it has been shown in agreement with \cite{Blaschke04}, that thermal instabilities arise in nNJL models. Three regulators, the Gaussian, integer and fractional Lorenzian regulators, have been analyzed and thermodynamic instabilities are present. It was shown that these instabilities are caused by the pressence of highly unstable poles of the propagator. There are two possible ways to handle this instabilities. On the one hand one can search for a parameter set that exhibits no such poles, or are otherwise highly suppressed, like in parameter set B for the Gaussian regulator. However, such a parameter set may not be possible. If this is the case then one is forced to carefuly select the poles that will be considered and leave the highly unstable poles out. This can be achieved by deforming the integration path in Fig. \ref{pathS} to surround the highly unstable poles. Although this may seem arbitrary there is a clear physical motivation for excluding these poles. Highly unstable poles have big decay widths and correspond to wide resonances. If we are to interpret these poles as quasiparticles then they are highly unstable quasiparticles that are widely spread. However, once we introduce the bosonic field $\sigma$ and take the mean field approximation, we are considering bound states of the quasiparticles. This highly unstable poles will not form physically well defined bound states since they represent very unstable quasiparticles. Furthermore, once we take the mean field value of the scalar field we are considering condensation of such quasiparticles. If one is to work in the mean field approximation then the model should consider well defined quasiparticles that can form bound states and condensate. In this manner, if the original effective propagator exhibits highly unstable poles, one should remove this poles in order to have well defined bosonic states.\\

In \cite{Blaschke04} it has been shown that the inclusion of the Polyakov loop in the model also contributes to soften these instabilities. In the next section it will be analyzed how the Polyakov loop affects the poles of the propagator for the Gaussian regulator case. Particularly, how it affects the pole with negative real part and, based on this analysis, why it contributes to soften the instabilities.\\

\section{The Polyakov loop and softened instabilities.}

The Polyakov loop $\langle\Phi\rangle$ is defined as \cite{Weise01}
\begin{equation}\langle\Phi(\boldsymbol{x})\rangle=\frac{1}{N_c}\langle\tr_c[L(\boldsymbol{x})]\rangle,\end{equation}
where $\tr_c$ is a trace over color indices and
\begin{equation}L(\boldsymbol{x})=\mathcal{P}\exp\left[i\int_0^\beta d\tau A_4(\tau, \boldsymbol{x})\right].\end{equation}
Here, $\mathcal{P}$ is the path-ordering operator and $A_4$ is the fourth component, in Euclidean space, of the gluon fields. The Polyakov loop can be incorporated into the model through the substitution $p_\mu\rightarrow p_\mu+A_\mu$ \cite{Weise01}. The Polyakov gauge \cite{Polyakov01} is considered, where only $A_\mu^3$ and $A_\mu^8$ are nonvanishing and, as in \cite{Weise02, Weise03, Blaschke06, Abuki01, Abuki02}, $A_\mu^8=0$. Then, in Minkoski space
\begin{equation}A_\mu=i\frac{\lambda_3}{2}A_0^3\delta_{\mu0}\equiv i\frac{\lambda_3}{2}\phi\delta_{\mu0},\end{equation}
where $\lambda_3$ is the third Gell-Mann matrix in $SU(3)$ color space. The propagator is then
\begin{equation}S(p)=(\slashed{p}-\Sigma(q))^{-1}\end{equation}
where $p_0=q_0+\frac{i\phi}{2}\lambda_3$ and $\boldsymbol{p}=\boldsymbol{q}$, $q$ being the four-momentum in Minkowski space. To look at the poles of the propagator, it can be rewritten like
\begin{equation}S(p)=(\slashed{p}+\Sigma(q))(\slashed{p}+\Sigma(q))^{-1}(\slashed{p}-\Sigma(q))^{-1}.\end{equation}
Here $(\slashed{p}+\Sigma(q))^{-1}(\slashed{p}-\Sigma(q))^{-1}$ is a matrix in Lorentz and color space. This matrix can be inverted and the propagator rewritten as
\begin{multline}S(\phi,q)=\\\frac{\slashed{p}+\Sigma}{(q^2-\Sigma^2(q))\left[(q^2-\Sigma^2(q)-\phi^2/4)^2+q_0^2\phi^2\right]}\boldsymbol{K}\label{propol},\end{multline}
where $\boldsymbol{K}=\boldsymbol{I}\otimes\boldsymbol{L}$. Here, $\boldsymbol{I}$ is the identity matrix in Lorentz space and $\boldsymbol{L}$ is a matrix in color space with no singularities
\begin{multline}\boldsymbol{L}=\mbox{diag}\left((q^2-\Sigma^2(q))\left(q^2-\Sigma^2(q)-iq_0\phi-\frac{\phi^2}{4}\right)\right.,\\(q^2-\Sigma^2(q))\left(q^2-\Sigma^2(q)+iq_0\phi-\frac{\phi^2}{4}\right),\\\left.\left|q^2-\Sigma^2(q)+iq_0\phi-\frac{\phi^2}{4}\right|^2\right)\end{multline}
From Eq. (\ref{propol}) it is clear that the usual poles we had at $q^2-\Sigma^2(q)=0$ are still there, but new poles have been added through the Polyakov loop. Such poles are solutions to
\begin{multline}(q^2-\Sigma^2-\phi^2/4)^2+q_0^2\phi^2=(q^2-\Sigma^2(q)-\phi^2/4)^2\\+(q^\mu A^3_{\mu})^2=0.\end{multline} 
This last equation is explicitly Lorentz covariant, so poles can be searched for in a reference frame where $\boldsymbol{q}=0$. In this manner, the new poles that have been included through the Polyakov loop can be found by solving $(q_0^2-\Sigma^2-\phi^2/4)^2+q_0^2\phi^2=0$. Of course, for $\phi=0$ we recover the poles we had before. This allows to study the behavior of a single pole as a function of the Polyakov loop.\\

In reference \cite{Blaschke04} the Gaussian regulator for parameter set A was shown to have softened the instabilities once the Polyakov look is included. As was discussed previously in this article, the main instability present in this case was caused by a negative real part pole. In this manner, since the instability is softened, the Polyakov loop should have an stabilizing effect on this pole.

\begin{figure}[!htb]
\begin{center}
\includegraphics[scale=0.25]{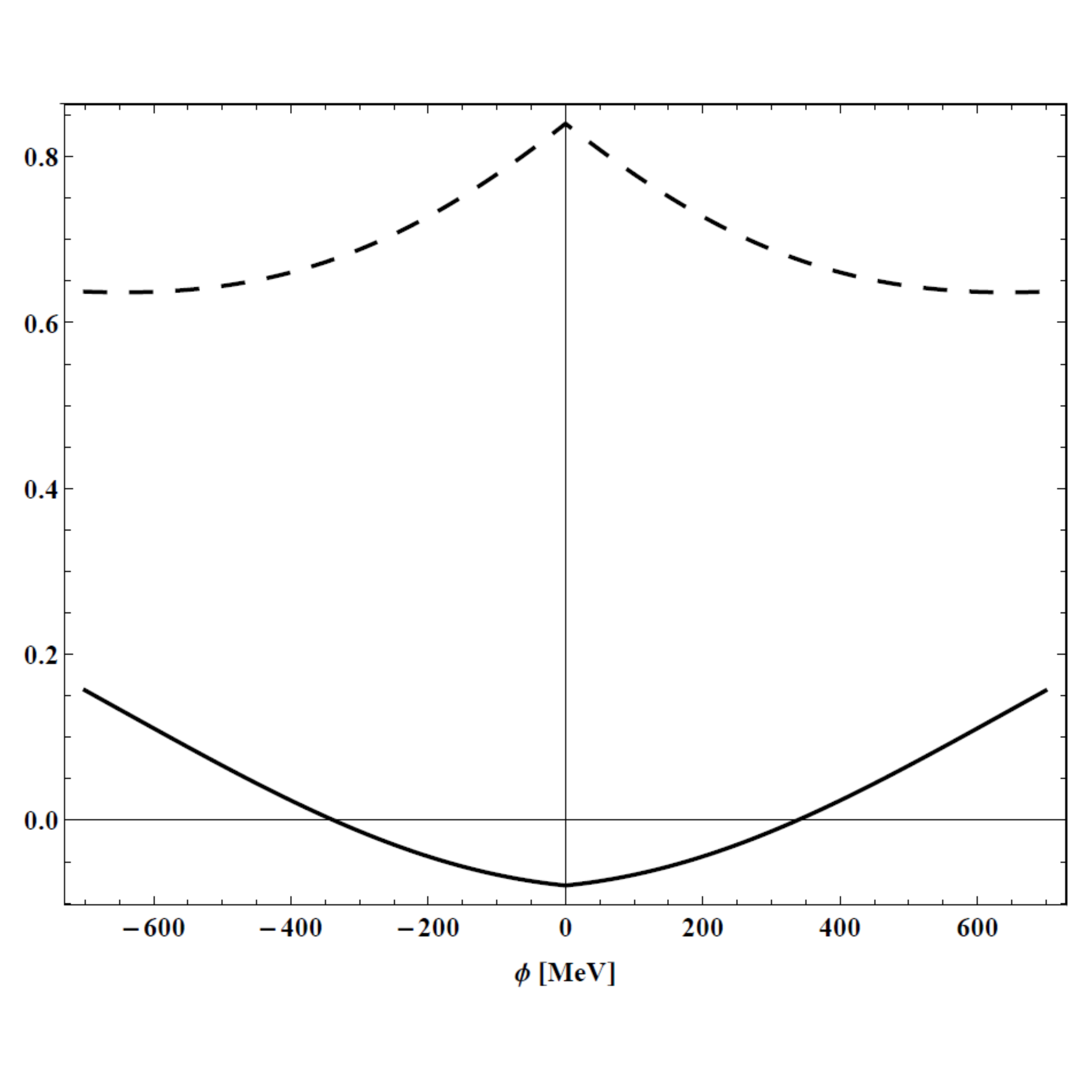}
\end{center}
\caption{Behavior of the negative real part pole from set A of the Gaussian regulator as a function of the Polyakov loop. The solid line is the real part of the pole and the dashed line is the imaginary part. Computations were made at $\bar{\sigma}=\bar{\sigma}_0$}
\label{f10}
\end{figure}

Figure \ref{f10} shows the evolution of the pole with a negative real part as a function of the Polyakov loop. As can be seen, the real part of the pole becomes positive at $\phi=\pm338$ MeV. Also, the imaginary part becomes smaller. It is clear that if $\phi=0$ MeV, then the usual poles are recovered and the instability would remain. However, once $\phi$ shifts to nonvanishing values the poles divide themselves. The old poles will still be there but new poles will also appear. If we focus in the negative real part pole, responsible for the instability, at nonvanishing values of the Polyakov loop, this pole will remain but a new pole will appear that no longer has a negative real part. When $\phi=0$ MeV both factors in the denominator of Eq. (\ref{propol}) contribute to the appearance of the negative real part pole. On the other hand, once $\phi$ is nonvanishing only one of those factor will produce the negative real part pole while the other will give rise to the new pole in Fig. \ref{f10}. In this manner, the contribution of the negative real part pole should be reduced in the case of nonvanishing Polyakov loop and hence, the instability will be softened. This is the reason behind the softening of instabilities found in reference \cite{Blaschke04} for the Gaussian regulator.\\

\section{Conclusions.}

The appearance of thermodynamic instabilities in the thermal nonlocal NJL model for three different regulators has been studied. For all three regulators it was shown that the instabilities were caused by the pressence of highly unstable poles of the propagator, i.e. poles with negative real parts or big imaginary parts. The removal of this poles eliminates the instability. On the other hand, well defined quasiparticles, i.e. real poles and poles with small imaginary parts, contribute to the expected behavior from a condensate and instabilities are not present when only these poles are considered.\\

The highly unstable poles do not represent well defined quasiparticles because of their big decay widths. Furthermore, since we introduce bosonic fields, we consider bound states of these poles. However these highly unstable poles will not form well defined bound states. In this manner, when working in the mean field approximation one should remove this highly unstable poles by deforming the integration path in Fig. \ref{pathS}. Also, in some cases, the instability may be avoided by searching for a set of parameters for which there are no highly unstable poles of the propagator.\\

It was also shown how the softened instabilities in the Gaussian regulator with Polyakov loop, may be understood by the effect the Polyakov loop has on the poles. The contribution of the negative real part pole in this case is reduced by the pressence of the Polyakov loop and a new pole is introduced. It is this fact that contributes to soften the instabilities.\\

\section{Acknowledgements.}

The author would like to thank M. Loewe and C. Villavicencio for helpful discussion. The author would like to acknowledge support from FONDECYT under grant No. 1130056 and CONICYT under Grant No. 21110577.

\bibliography{Insta}{}
\bibliographystyle{apsrev4-1}

\end{document}